\documentclass[sigconf,nonacm]{acmart}

\usepackage{amsmath, amssymb, algorithm, algorithmic}
\usepackage{enumitem}
\usepackage{subcaption}

\AtBeginDocument{%
  }
\begin{document}

\title{MaRCA: Multi-Agent Reinforcement Learning for Dynamic Computation Allocation in Large-Scale Recommender Systems}

\author{Wan Jiang}
\email{jiangwan1@jd.com}
\affiliation{%
  \institution{JD.com}
  \city{Beijing}
  \country{China}
}

\author{Xinyi Zang}
\email{zangxinyi3@jd.com}
\affiliation{%
  \institution{JD.com}
  \city{Beijing}
  \country{China}
}

\author{Yudong Zhao}
\email{zhaoyudong10@jd.com}
\affiliation{%
  \institution{JD.com}
  \city{Beijing}
  \country{China}
}

\author{Yusi Zou}
\email{zouyusi1@jd.com}
\affiliation{%
  \institution{JD.com}
  \city{Beijing}
  \country{China}
}

\author{Yunfei Lu}
\email{luyunfei1@jd.com}
\affiliation{%
  \institution{JD.com}
  \city{Beijing}
  \country{China}
}

\author{Junbo Tong}
\email{tjb21@mails.tsinghua.edu.cn}
\affiliation{%
  \institution{Tsinghua University}
  \city{Beijing}
  \country{China}
}

\author{Yang Liu}
\email{liuyang123@jd.com}
\affiliation{%
  \institution{JD.com}
  \city{Beijing}
  \country{China}
}

\author{Ming Li}
\email{liming666@jd.com}
\affiliation{%
  \institution{JD.com}
  \city{Beijing}
  \country{China}
}

\author{Jiani Shi}
\email{shijiani@jd.com}
\affiliation{%
  \institution{JD.com}
  \city{Beijing}
  \country{China}
}


\author{Xin Yang}
\email{yangxin81@jd.com}
\affiliation{
  \institution{JD.com}
  \city{Beijing}
  \country{China}
}

\renewcommand{\shortauthors}{Wan Jiang et al.}

\begin{abstract}
Modern recommender systems face significant computational challenges due to growing model complexity and traffic scale, making efficient computation allocation critical for maximizing business revenue. Existing approaches typically simplify multi-stage computation resource allocation, neglecting inter-stage dependencies, thus limiting global optimality. In this paper, we propose \textbf{MaRCA}, a \textbf{m}ulti-\textbf{a}gent \textbf{r}einforcement learning framework for end-to-end \textbf{c}omputation resource \textbf{a}llocation in large-scale recommender systems. MaRCA models the stages of a recommender system as cooperative agents, using Centralized Training with Decentralized Execution (CTDE) to optimize revenue under computation resource constraints. We introduce an AutoBucket TestBench for accurate computation cost estimation, and a Model Predictive Control (MPC)-based Revenue-Cost Balancer to proactively forecast traffic loads and adjust the revenue‑cost trade‑off accordingly. Since its end-to-end deployment in the advertising pipeline of a leading global e-commerce platform in November 2024, MaRCA has consistently handled hundreds of billions of ad requests per day and has delivered a 16.67\% revenue uplift using existing computation resources.
\end{abstract}




\maketitle

\section{Introduction}
\label{sec:introduction}

Modern recommender systems analyze user behavior and contextual information to filter relevant items from large candidate pools and generate a ranked list of recommendations through multi-stage processing pipelines \cite{cheng2016widedeeplearning, DIN}. These systems have become crucial infrastructure for e-commerce platforms \cite{ecommerce}.

Contemporary industrial recommender systems typically adopt a cascaded architecture comprising three stages: retrieval, pre-ranking, and ranking \cite{copr, Liu_2017}. Each stage of recommender systems involves models of different complexity and is subject to specific computational constraints.  However, most early academic research on recommender systems aims to maximize profit under the assumption of abundant computation resources \cite{prior1, prior2, prior3}, failing to address resource allocation constraints.

\begin{figure}[h!]
  \centering
  \includegraphics[width=0.45\textwidth]{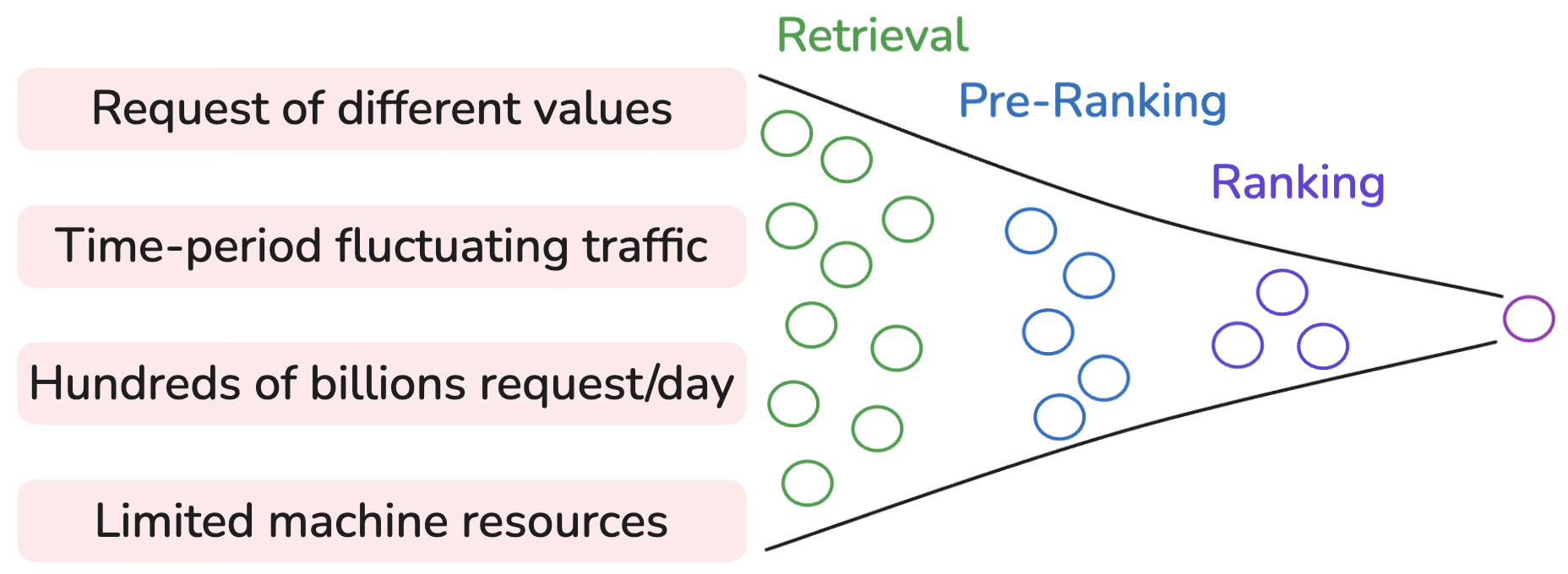} 
  \caption{Overview of the general architecture of recommender systems, highlighting the key stages and components involved in processing user requests.}
   \Description[Alt Text]{Multi-stage recommender pipeline processing user requests from retrieval to ranking stages.}
  \label{fig:backgroundfig}
\end{figure}

With the continued advancement of deep learning-driven recommender systems, the tension between computational demand and available machine resources has become increasingly significant. In industrial recommender systems, traffic volume varies significantly across different periods \cite{temporal_traffic}, and the value of requests differs across media platforms and user demographics \cite{request_value}. An effective computation allocation strategy should dynamically adapt to fluctuating traffic conditions while maintaining system stability and high recommendation quality within limited machine resources. Figure~\ref{fig:backgroundfig} illustrates an overview of such a multi-stage recommender system.

To address these issues, recent efforts in Dynamic Computation Allocation (DCA) have explored two major categories of approaches: optimization-based methods and reinforcement learning (RL)-based methods. Optimization-based methods, such as linear programming and heuristic scheduling, allocate computation resources based on predefined constraints and business rules. These methods offer fast and interpretable decision-making but often struggle to adapt to dynamic traffic fluctuations and evolving system conditions. In contrast, RL-based methods frame resource allocation as a sequential decision-making problem, enabling flexible strategies by learning from historical data and real-time interactions. However, many reinforcement learning methods have adopted either a single-agent paradigm or centralized multi-agent architectures, which tend to overlook the localized operational constraints inherent in distributed recommender systems.



In industrial settings, key stages of the recommendation pipeline are deployed across different data centers, each operating under distinct constraints and relying on localized observations. Centralized approaches combine all decisions into a single prediction, which may overlook localized factors. To overcome these constraints, we propose MaRCA, a cooperative multi-agent framework that models each stage as an autonomous agent. Using Centralized Training with Decentralized Execution (CTDE), each agent makes decisions using local information while benefiting from coordinated global learning. This approach leverages cooperative agent collaboration to maximize business revenue under limited computation resources. Extensive offline experiments of MaRCA 
and a four-week online A/B test on a major e-commerce platform demonstrate that MaRCA achieves a \textbf{16.67\% increase in advertising revenue} without additional computation cost (see more details in Section~\ref{experiments}).

Our main contributions are summarized as follows:
\begin{enumerate}
    \item {We propose a novel collaborative multi-agent framework that leverages the \textbf{Adaptive Weighting Recurrent Q-Mixer (AWRQ-Mixer)}, which integrates an adaptive weighting recurrent Q for sequential decision-making with a mixing network to foster cross-stage coordination. By employing CTDE, our framework achieves globally optimal resource allocation.}
    \item {We develop an \textbf{MPC-based Revenue-Cost Balancer} that optimally allocates resources by forecasting traffic fluctuations. This approach enables real-time, proactive decision-making, and therefore enhances system stability and efficiency under dynamic resource constraints.}
    \item {We design an \textbf{AutoBucket TestBench} for computation cost estimation (see more details in Appendix \ref{sec:autobucket}). Our framework employs automated testing and intelligent bucketing to address the absence of explicit cost labels.}
\end{enumerate}

\section{Related Work}

Dynamic computation allocation (DCA) in large-scale recommender systems has been studied under paradigms ranging from deterministic optimization to learning-based methods.

\subsection{Deterministic Optimizers for Computation Allocation}

Early solutions to manage computational load relied on heuristic-based elastic degradation or static rules (e.g., disabling heavy models or truncating result lists) \cite{google_sre, autoscaling}, while robust, they are value-agnostic and struggle with dynamic traffic. 

The first value-aware solutions formulate DCA as a constrained optimization problem. DCAF \cite{DCAF} casts resource allocation as a 0–1 knapsack that maximizes revenue under a global computation budget. CRAS \cite{CRAS} augments this idea with a PID feedback controller \cite{PID}, improving stability during bursts. SACA \cite{SACA} simultaneously employs an elastic queue and elastic model, enabling the incorporation of different action types within a single module. However, its binary-search tuning lags behind rapid traffic shifts. GreenFlow \cite{greenflow} is a learning-augmented deterministic optimizer that combines reward prediction with a dynamic primal–dual solver to allocate multi-stage action chains under a global computation budget.

These works validate the feasibility of DCA but were founded on a key simplification that recommendation stages are independent and that their costs are static. Meanwhile, purely reactive control can lead to oscillations. These limit their ability to capture the complex, non-stationary inter-stage dependencies of a live recommender system.

\subsection{Learning-Based Approaches for Computation Allocation}

To overcome the limitations of deterministic optimizers, researchers recast DCA as a sequential decision process and applied reinforcement learning (RL). Classic RL variants such as DQN \cite{DQN}, DDQN \cite{DDQN}, and DRQN \cite{DRQN} address high-dimensional, partially observable environments. Averaged Ensemble-DQN \cite{averagedDQN} improves the accuracy of Q-value estimation by averaging the outputs of multiple Q-networks, while REM \cite{REM} improves generalization via random value mixing, at the cost of additional noise.

A representative learning-based baseline of DCA is RL-MPCA \cite{RL-MPCA}, which treats the multi-stage pipeline as a weakly coupled MDP \cite{wcmdp} and lets a single agent coordinate all stages. While representing a significant advance over deterministic methods, centralized single-agent methods do not align well with the decentralized reality of industrial recommender systems, where stages run on separate services \cite{copr, youtube_dnn}. 

These considerations naturally lead to multi-agent reinforcement learning (MARL) formulations \cite{marl-survey, marl-survey2, marl-survey3}. The Centralized Training with Decentralized Execution (CTDE) paradigm \cite{CTDE} allows agents to learn a globally coordinated policy while executing actions based on local observations. Within the CTDE paradigm, MARL methods split into centralized-critic and value-decomposition families \cite{RAO}. Centralized-critic algorithms \cite{MADDPG, MAPPO} must model an exponential joint-action space and ingest the full global state, limiting scalability in complex industry settings. Value-decomposition approaches, such as VDN \cite{VDN} and QMIX \cite{QMIX}, factorize the global value into per-agent utilities, achieving better scalability.

MARL has been applied to various areas in recommender systems, including multi-stage recommendation coordination \cite{DeepChain,MARIS,UNEX}, ad slot ranking \cite{content_ranking}, ad bidding \cite{DCMAB, DBLP}. Separately, RL-based resource scheduling has been extensively explored in infrastructure management \cite{rm1, rm2}. However, using MARL for resource allocation within recommender systems remains underexplored. To our knowledge, MaRCA is the first fully cooperative MARL framework for end-to-end computation allocation in a recommender system, bridging the gap between RL and industrial DCA.

\section{Methodology}
\label{sec:methodology}
\subsection{Problem Formulation}
\label{sec:constrainedOptimization}

To address the challenges outlined in the previous sections, we formulate the multi-stage recommendation process as a constrained sequential decision-making problem that aims to maximize overall business revenue while adhering to strict computation resource constraints. We formally define the following:
\begin{itemize}
    \item \textbf{State Space \(\mathcal{S}\).} At each step \(t\), the system observes a state \(s_t \in \mathcal S\), encapsulating user profile features, real-time traffic patterns, and resource utilization metrics. 
    \item \textbf{Action Space \(\mathcal{A}\).} At each step $t$, the joint action combination is $\mathbf{a}=(a_1,\dots,a_n)\in\mathcal{A}$. The joint action space is $\mathcal{A}=\mathcal{A}_1\times\dots\times\mathcal{A}_n$. The action combination \(\mathbf{a}\) collectively specifies the decisions across multiple stages, including retrieval channels to activate, switch modules to enable, and queue truncation lengths. 
    \item \textbf{Action Value $Q(s_t,a_t)$ .} Given the current state \(s_t\) and action \(a_t\), $Q(s_t,a_t)$ denotes the expected business revenue. 
    \item \textbf{Computation Cost $C(s_t,a_t)$}. $C(s_t,a_t)$ represents the computation resources required to execute the action \(a_t\) given the current state \(s_t\). We define:
    \begin{equation}
        C(s_t,a_t) = \hat{C}(s_t,a_t) + f(D_{t})
    \end{equation}
    where $\hat{C}(\cdot)$ is the computation cost predicted by the AutoBucket TestBench (see more details in Appendix \ref{sec:autobucket}) and $D_{t}$ is the elastic degradation level mapped into equivalent computation cost through $f(\cdot)$ (calibrated in isolated load tests).
      
    \item \textbf{Reward $R(s_t,\mathbf{a})$.} The reward function is designed as business revenue. Unlike conventional step-wise rewards, our system can only observe business revenue after the entire action sequence for a request is completed.
\end{itemize}

Consider a batch of \(M\) user requests indexed by \(i\). For a request \(i\), the system assigns a binary decision variable \(x_{i,\mathbf{a}} \in \{0,1\}\) indicating whether a specific action combination \(\mathbf{a}\) is selected (\(x_{i,\mathbf{a}} = 1\)) or not (\(x_{i,\mathbf{a}} = 0\)). 
We impose a computation resource budget \(C_{m}\) to limit the overall computation cost over all requests. Following \cite{DCAF}, we formulate the constrained optimization problem as stated in Eqs. \eqref{eq:obj}–\eqref{eq:constr3}. 

\begin{align}
\label{eq:obj}
\max_{{x_{i,\mathbf{a}}}} \quad 
& \sum_{i=1}^M \sum_{\mathbf{a} \in \mathcal{A}} x_{i, \mathbf{a}}\,Q(s_t,\mathbf{a}) \\[4pt]
\label{eq:constr1}
\text{s.t.} \quad 
& \sum_{i=1}^M \sum_{\mathbf{a} \in \mathcal{A}} x_{i,\mathbf{a}}\,C(s_t, \mathbf{a}) \le C_m \\[2pt]
\label{eq:constr2}
& \sum_{\mathbf{a} \in \mathcal{A}} x_{i,\mathbf{a}} = 1,\quad \forall\, i = 1,\dots, M \\[2pt]
\label{eq:constr3}
& x_{i,\mathbf{a}} \in \{0,1\},\quad \forall\, i = 1,\dots, M,\; \mathbf{a} \in \mathcal{A}
\end{align}

We enforce a one-hot structure over the joint action space. For each request \(i\), exactly one composite action is executed. 

To satisfy the budget constraint in Eqs. \eqref{eq:constr1}--\eqref{eq:constr3} while maximizing total revenue in Eq. \eqref{eq:obj}, we adopt a Lagrangian relaxation approach, as in \cite{DCAF}. The complete derivation is provided in Appendix~\ref{sec:lagrangian}. This yields the following request-level decision rule:

\begin{equation}
\label{eq:a_star}
\mathbf{a}^* = \arg\max_{\mathbf{a}\in\mathcal{A}} \left( Q(s_{t}, \mathbf{a}) - \lambda C(s_{t}, \mathbf{a}) \right)
\end{equation}

Here, \(\mathbf{a}^*\) represents the chosen action combination that maximizes the net benefit, measured by the revenue \(Q(s_{t}, \mathbf{a})\) minus the \(\lambda\)-weighted cost \(C(s_{t}, \mathbf{a})\). Through MaRCA, we can learn appropriate policy parameters and dynamically adapt \(\lambda\) based on real-time load.

\subsection{System Design}

\label{systemdesign}

\begin{figure}[ht]
  \centering
  \includegraphics[width=0.5\textwidth]{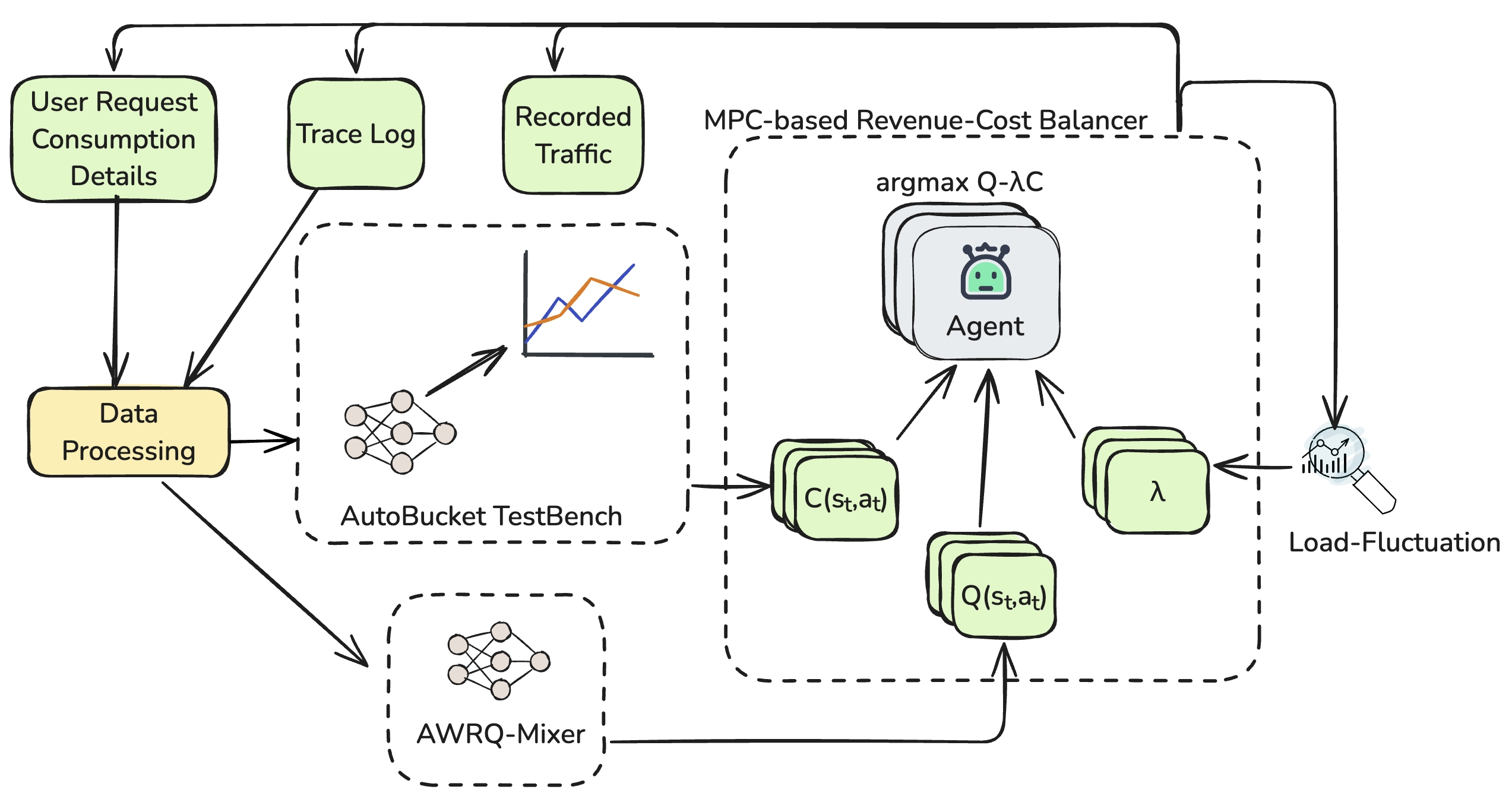} 
  \caption{MaRCA system architecture: multi-agent collaborative decision flow with Adaptive Weighting Recurrent Q-Mixer, AutoBucket TestBench, and MPC-Based Revenue-Cost Balancer.}
  \Description[Alt Text]{MaRCA system workflow with AWRQ-Mixer, AutoBucket TestBench, and MPC-based Revenue-Cost balancer.}
  \label{fig:systemflowfig}
\end{figure}

As illustrated in Figure~\ref{fig:systemflowfig}, the system follows a collaborative multi-agent framework, where the AWRQ-Mixer and AutoBucket TestBench feed their computed metrics into the MPC-based Balancer, which then orchestrates the final action selection. The AWRQ-Mixer assesses the expected business revenue \(Q(s_t, \mathbf{a})\) under various states and expected action combinations. Meanwhile, the AutoBucket TestBench (see more details in Appendix \ref{sec:autobucket}) processes trace-log data and predicted action outcomes to estimate the computation cost \(C(s_t, \mathbf{a})\) for each action combination. Subsequently, the MPC-based revenue-cost balancer dynamically selects optimal actions by balancing \(Q(s_t, \mathbf{a})\) and \(C(s_t, \mathbf{a})\), guided by real-time resource utilization. 

\subsection{Adaptive Weighting Recurrent Q-Mixer}
\label{sec:awrqmixer}
The action value estimation module, AWRQ-Mixer, predicts the expected revenue of each request by jointly encoding user attributes, contextual information, and the inter‑stage dependencies in the recommendation stages.


To illustrate why modeling these interdependencies matters, consider that the stages of a recommender pipeline are highly interdependent, with decisions made upstream directly constraining what can be achieved downstream. For example, changes in retrieval actions can alter the candidate pool and thus the final revenue, even when the ranking actions remain the same. However, due to the independent operation of these stages in separate service clusters with resource constraints \cite{RL-MPCA} \cite{twerc}, it is necessary to model their interdependencies while maintaining their ability to independently manage computation costs.

AWRQ-Mixer meets this requirement through three innovations: (1) an adaptive weighting recurrent Q ensemble that dynamically integrates multiple Q-value estimators; (2) a variance-guided credit assignment mechanism to allocate reward among actions; and (3) a softplus-based monotonicity constraint that ensures cooperative aggregation of agent-level values. The following subsections will detail each of these innovations.

\subsubsection{Adaptive Weighting Recurrent Q (AWRQ)} 
Traditional DQN performs well in fully observable environments but is limited in partially observable environments due to incomplete state information. In such contexts, historical observations become crucial for accurately inferring the underlying state. Therefore, we employ DRQN to process sequences of observations over time:
\begin{equation}
    h_t = \mathrm{GRU}\bigl(o_t, h_{t-1}\bigr), \quad Q(o_t,a_t) = \mathrm{MLP}(h_t)
\end{equation}
where \(o_t\) is the observation, and \(h_t\) is the hidden state capturing historical context.

However, a single Q-value estimator in DRQN is insufficient to capture the multi‑faceted, cross‑stage decision process. Inspired by ensemble learning principles, we extend DRQN by introducing parallel recurrent Q-value estimators to address uncertainty in the estimation process. Each agent instantiates a recurrent ensemble of $K$ heads. For each head $k\in\{1,\dots,K\}$, it outputs $Q^{k}_{\theta_g}(s_t, a_t)$, where $g\in\{1,\dots,n\}$ indexes the agents. For brevity, $Q_{\theta_g}$ is henceforth written as $Q_g$ whenever parameters are clear from context. Rather than averaging, we dynamically weight ensemble outputs according to their temporal difference (TD) errors, and we call this method Adaptive Weighting (AW).

At each training step, each Q-head's TD error is recorded. If a head exhibits a larger error on a given mini‑batch, it is assigned a correspondingly larger weight among the heads. This design intentionally emphasises under‑performing heads, ensuring they receive greater focus during training and can be corrected more quickly. Empirically, this weighting scheme accelerates convergence and enhances robustness.

Consider a mini-batch in which each sample contains the state $s_t$, the action $a_t$, and the next state $s_{t+1}$. The loss for each Q-head is then computed as follows:
 
\begin{equation}
\label{eq:aw-loss}
    \mathcal{L}_{k,t} = \left( r_t + \gamma Q_{g'}^k(o_{t+1},  a_{t+1}) - Q_{g}^k(o_t, a_t) \right)^2
\end{equation}

where $\gamma$ is the discount factor, and $Q_{g'}(o_{t+1}, a_{t+1})$ is the target Q-value for the next state-action pair, calculated by the target network.

The individual losses are normalized to compute the adaptive weight $\eta_{k,t}$ for each Q-head:
  
\begin{equation}
\label{eq:aw}
\eta_{k,t} = \frac{\mathcal L_{k,t}}{\sum_{k=1}^K \mathcal L_{k,t}}
\end{equation}

Finally, the agent’s Q-value is taken as the weighted sum:

\begin{equation}
Q_{g} = \sum_{k=1}^K \eta_{k,t} \, Q_{g}^k(o_t, a_t)
\end{equation}

\subsubsection{Softplus-Based Monotonicity Constraints (SMC) for Cooperative Agents}
In multi-agent recommendation pipelines, the joint action-value \(Q_{\mathrm{tot}}\) must be non-decreasing in each agent’s value \(Q_g\) to reflect their cooperative contribution. To enforce this, we define a mixing network $\mathcal{M}$ that aggregates individual Q-values \( Q_{{1}}, Q_{{2}}, \ldots, Q_{{n}} \) from multiple agents into a joint Q-value \( Q_{\text{tot}} \):

\begin{equation}
Q_{{tot}} = \mathcal{M}( Q_{{1}}, Q_{{2}}, \ldots, Q_{{n}} , s_t)
\end{equation}

The parameters of the mixing network are generated by a hypernetwork \(h_{\psi}\) by taking the state $s_t$ as input:

\begin{equation}
(\, \tilde W_1,\,b_1,\, \tilde W_2,\,b_2\,)
\;=\;
h_{\psi}(s_t)
\end{equation}

Because the stages cooperate, the joint value must be monotonic non-decreasing in every agent's value. We enforce this by applying the Softplus transform to ensure non-negativity of all weight matrices and thus ensure \( Q_{\text{tot}} \) is a monotonic function of each \( Q_{{g}} \). 

\begin{equation}
\text{Softplus}(W) = \ln(1 + e^{W})
\end{equation}

\begin{equation}
\label{eq:smc}
W_i = \mathrm{Softplus}(\tilde W_i),\quad i\in\{1,2\}
\end{equation}

The mixing network is trained by minimizing
\begin{equation}
\mathcal L\left({\theta_{\text {tot }}}\right)=\left(r_t+\gamma Q_{{\text {tot }}^{\prime}}\left(s_{t+1}, \mathbf{a}^{\prime}\right)-Q_{{\text {tot}}}\left(s_t, \mathbf{a}\right)\right)^2
\end{equation}

\subsubsection{Variance-Guided Credit Assignment (\textbf{VGCA})}
In environments with sparse or delayed rewards, such as large-scale recommendation systems, it's crucial to determine how each action contributed to the final reward. To address this, we introduce an auxiliary reward signal that mitigates sparse rewards and enhances training stability. The key insight is using variance across candidate actions to guide their contribution to the final reward. 

        


\begin{equation}
\label{eq:vgca-weight}
w_t = \mathbb{V}\mathrm{ar}_{j\in\mathcal{A}_t}\!\bigl[\,\mathbb{E}[\,R\mid a_t=j\,]\,\bigr]\ 
\end{equation}

where $r_t$ is the reward and $\mathcal{A}_t$ is the discrete action space of $a_t$. During TD updates, the reward $r_t$ of each action is scaled by $w_t$, amplifying learning signals for high-impact dimensions. This adjustment ensures that agents whose actions induce greater variance, and therefore have a greater potential impact on revenue, receive proportionally stronger learning signals.  This auxiliary signal serves as an additional guide to the model during training, helping it better identify and prioritize the most relevant actions, even in the absence of frequent immediate feedback. 
Algorithm~\ref{alg:awrqmixer-training} summarizes the overall training procedure for AWRQ-Mixer.

\begin{figure*}[ht]
  \centering
  \includegraphics[width=0.9\textwidth]{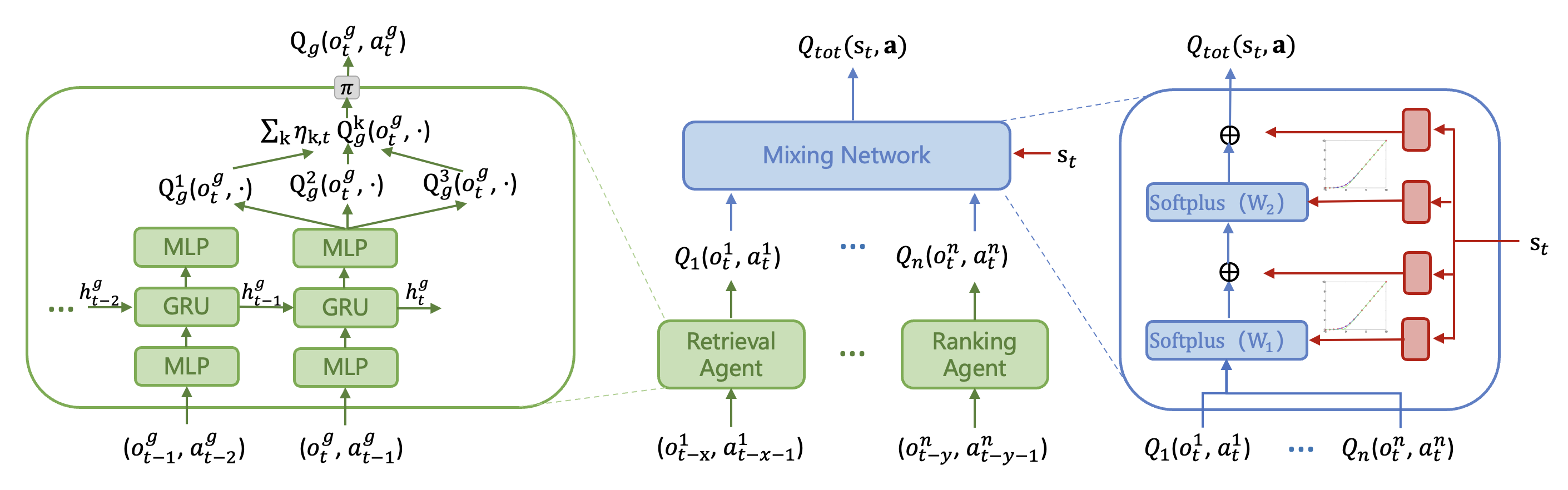} 
  \caption{Adaptive Weighting Recurrent Q-Mixer (AWRQ-Mixer) Architecture in MaRCA.}
  \Description[Alt Text]{Detailed schematic of the Adaptive Weighting Recurrent Q-Mixer (AWRQ-Mixer), illustrating the cooperative agents and mixing network structure.}
  \label{fig:action_value_model}
\end{figure*}

Bringing these elements together, AWRQ-Mixer extends a DRQN backbone with (1) Adaptive Weighting Recurrent Q, (2) Variance-Guided Credit Assignment, and (3) Softplus-Based Monotonicity Constraints. As illustrated in Figure~\ref{fig:action_value_model}, the framework models the recommendation stages as cooperative agents through AWRQ. The mixing network aggregates  Q-values from AWRQ while enforcing monotonicity, guided by state information $s_t$ to dynamically tailor mixing weights using a hypernet. At training time, we adopt centralized optimization of all agents, allowing us to capture global dependencies. At inference time, each agent can operate independently in a separate cluster, thus enabling scalable decentralized execution without extra cross-agent communication overhead. This design is particularly critical in production environments where stages such as retrieval and ranking often run in physically separate machine clusters. 

\begin{algorithm}[ht]
\caption{Offline Training of AWRQ-Mixer}
\label{alg:awrqmixer-training}
\begin{algorithmic}[1]
\REQUIRE Dataset $\mathcal{D}$, \#iterations $I$, mini-batch size $N$, agent set $\mathcal{G}$, ensemble heads $K$, discount $\gamma$, hypernet $h_{\psi}$, target update frequency $\tau$.
\STATE Initialize AWRQ $\{Q_{g}\}_{g\in\mathcal G}$ (each with $K$ heads) and targets $Q_{g'}$ for each agent, initialize mixing network hypernet $h_{\psi}$.
\FOR{$\textit{iter}=1$ \textbf{to} $I$}
    \STATE Sample a mini-batch $\mathcal{B}$ of transitions $\{(s_t, o_t,a_t, R, s_{t+1})\}$ from $\mathcal{D}$.
    \FOR{t=1,\dots,T}
        \STATE  $\forall t$: $w_t = \mathbb{V}\mathrm{ar}_{j\in\mathcal{A}_t}\!\bigl[\,\mathbb{E}[\,R\mid a_t=j\,]\,\bigr]\ $ \hfill // VGCA, \textbf{Eq. \eqref{eq:vgca-weight}}
        \STATE $r_t = w_t\, R$
    \ENDFOR
    \FOR{$t=1$ \textbf{to} $T$}
        \STATE AWRQ outputs $K$ Q-values $ Q_{g}^k(o_t, a_t)$.
        \STATE {\raggedright From \textbf{Eq. \eqref{eq:aw-loss}} and \textbf{Eq. \eqref{eq:vgca-weight}}:\par}
        \(
          \begin{aligned}[t]
            \ \ \ \ &\mathcal L_{k,t}
              = \bigl(r_t
                   + \gamma\,Q_{g'}^k(o_{t+1},a_{t+1})\bigr)
              -\;Q_{g}^k(o_t,a_t)
            \bigr)^2
          \end{aligned}
        \)
        \STATE   \(\eta_{k,t} = \mathcal L_{k,t} / \sum_{k=1}^K \mathcal L_{k,t}\)    \hfill // AW, \textbf{Eq. \eqref{eq:aw}}
        \STATE \(Q_{g} = \sum_{k=1}^K \eta_{k,t} \, Q_{g}^k(o_t,  a_t)\)
        \STATE  $ \theta_{g} \leftarrow \operatorname*{arg\,min}_{\theta_{\text{g}}} \mathcal L_{k,t} $
        \STATE $\theta_g' \gets \theta_g $ if $\mathrm{iter}\bmod \tau=0$
    \ENDFOR
\STATE  $(W_1,b_1,W_2,b_2)\leftarrow h_{\psi}\!\big(s_t,\mathbf  a\big)$ 
with $W_1=\mathrm{softplus}(\cdot)$, $W_2=\mathrm{softplus}(\cdot)$ \hfill // SMC, \textbf{Eq. \eqref{eq:smc}} 
\STATE $Q_{\mathrm{tot}}
  = \mathrm{ReLU}\!\bigl([Q_{{1}},\dots,Q_{{n}}]\,W_1+b_1\bigr)\,W_2+b_2$
\STATE \(\mathcal L_{\text{tot}}= \left(R + \gamma Q_{{\text{tot}}'}(s_{t+1}, \mathbf a') - Q_{{\text{tot}}}(s_t, \mathbf a)\right)^2\)
\STATE \(\psi \leftarrow \operatorname*{arg\,min}_{\psi} \mathcal L_{\text{tot}}\)
\ENDFOR
\ENSURE Trained $\{\theta_g\}$ and $\psi$.
\end{algorithmic}
\end{algorithm}

\subsection{MPC-Based Revenue-Cost Balancer} 
\label{sec:asmpc}
The MPC-Based Revenue-Cost Balancer determines the optimal \( \mathbf a^* \) that maximizes the expected reward in a given state \( s \) while keeping the computation resource utilization within the computation budget \( C_{m} \). In prior works such as DCAF \cite{DCAF} and SACA \cite{SACA}, binary search is employed to determine $\lambda$. Moreover, CRAS \cite{CRAS} and RL-MPCA \cite{RL-MPCA} utilize feedback-based methods to adjust $\lambda$ for dynamic adaptation.

However, such feedback-based control inherently incurs one-step latency in responding to traffic fluctuations, leading to oscillatory resource misallocation when compensating for sudden traffic changes. To address this, we design a Model Predictive Control (MPC) \cite{mpc} framework that performs rolling-horizon optimization. By continuously predicting incoming traffic patterns and precomputing optimal $\lambda$ trajectories, our approach enables proactive stabilization.

MPC optimizes computation allocation over a finite time horizon \( N \) by continuously solving an optimization problem that balances business performance, latency constraints, and hardware efficiency. The optimization process aims to ensure that the system computation resource utilization remains close to the computation budget and minimizes fluctuations to maintain stability. 

The optimization goal at time \( t \) is defined as:

\begin{equation}
\label{eq:mpc_object_refined}
\begin{aligned}
J = \min_{\substack{\lambda_{t+i} \\ 0 \le i \le N-1}} \;\;
& \sum_{i=0}^{N} \alpha^i \left\|\hat{C}_{t+i} - C_m\right\|^2 \\
& + 
\begin{cases}
0, & \text{if } \hat{C}_{t+i} < C_m, \\
\beta \sum_{i=0}^{N} \alpha^i 
\left\|\hat{C}_{t+i} - \hat{C}_{t+i-1}\right\|^2, & \text{otherwise.}
\end{cases}
\end{aligned}
\end{equation}

\begin{equation}
\begin{aligned}
\text{s.t.} \quad  \hat{C}_{t+i+1} &= g(\hat{C}_{t+i}, s_{t+i}, \lambda_{t+i}) \\
0 &\leq i \leq N-1
\end{aligned}
\end{equation}

Here \(\alpha^i\in (0,1]\) is a decay weighting factor that reduces the weight of longer-term predictions, thus mitigating long-horizon errors. \(\beta\) is an oscillation damping factor penalizing abrupt computation resource utilization changes.
\(g(\cdot)\) is a learned system model mapping computation resource states \(\hat{C}_{t+i}\), environment states \(s_{t+i}\), and the revenue-cost balancer \(\lambda_{t+i}\) to future CPU loads.

By optimizing \(J\), this objective ensures that computation resource utilization near \(C_m\) while reducing fluctuations. Only the first element of the computed optimal sequence \( \{\lambda_{t}^*, ..., \lambda_{t+N-1}^*\} \) is applied at time \( t \), ensuring proactive and stable resource allocation.

\section{Experiments}
\label{experiments}

\subsection{Offline Experiments}
Our evaluation focuses on two core modules: the AWRQ-Mixer and the MPC-based Revenue–Cost Balancer. We evaluate the AWRQ-Mixer using both model metrics and simulated revenue derived from real-world logs, demonstrating its prediction accuracy and revenue uplift. For the MPC-based Balancer, we introduce utilization and overutilization rates as evaluation metrics to quantify its scheduling capability.

\subsubsection{AWRQ-Mixer Experiment Results}

\paragraph{\textbf{Implementation details}}  The AWRQ-Mixer relies on several hyperparameters. We used grid search \cite{grid} to identify the optimal hyperparameter values. A list of these parameters and their values is provided in Appendix~\ref{sec:hyperparameters}.

\paragraph{\textbf{Model Evaluation Metrics}} To quantify the monotonic relationship between the model’s predicted rankings and the true rankings, we use Spearman's Rank Correlation ($r_s$):
    \begin{equation}
    r_s = \frac{\mathrm{cov}(\mathrm{R}(X), \mathrm{R}(Y))}{\sigma_{\mathrm{R}(X)} \sigma_{\mathrm{R}(Y)}}
    \end{equation}
    where \(\mathrm{cov}(\mathrm{R}(X), \mathrm{R}(Y))\) denotes the covariance between the ranks of variables \( X \) and \( Y \), and $\sigma_{\mathrm{rank}(X)}$, $\sigma_{\mathrm{rank}(Y)}$ are the corresponding standard deviations. Higher values of \( r_s \) indicate stronger positive correlations, reflecting high agreement between the predicted and actual orderings.  

 \paragraph{\textbf{Training Stability Metrics}} To quantify optimisation stability we report two extra metrics:

 \begin{itemize}
\item Convergence. The number of environment steps (in millions) required for a model to reach 95\% of its final validation Return\% (averaged over 5 random seeds). A smaller value indicates faster sample efficiency.

\item Gradient‑variance. $g_t=\lVert\nabla_\theta\mathcal{L}_t\rVert_2$ be the L2‑norm of the joint‑Q network gradient. For a sliding window of 1000 updates, we compute the variance and then average across all windows and seeds. Lower variance implies smoother gradient flow and more stable learning.
\end{itemize}

\paragraph{\textbf{Revenue Simulation}}
Evaluating new models in a live environment is often risky and resource-intensive. To enable fair and controlled comparisons under consistent computation budgets, we simulate revenue through four steps:

\begin{itemize}
    \item Ground Truth Estimation. Train an ensemble model on historical data, including train and test data, to approximate true revenue (with \(r_s\) = 0.95). 

    \item Uniform Computation Cost. Derive action distributions from the test data that reflect the total computation cost, and use them as the action quota.

    \item Action Allocation. Train the baseline models on the training dataset, and compute the action values \(\hat{Q}(s,\mathbf{a})\) for all action-state pairs in the test dataset. Sort all \((s,\mathbf{a})\) pairs by \(\hat{Q}\), then iteratively assign the highest-valued action without exceeding the action quota. 

     \item Revenue Evaluation. Calculate total expected revenue using the ensemble model and assess performance using Relative Return Percentage:
    \begin{equation}
        \text{Return\%} = \frac{\sum Q_{\text{model}}(s,a)}{\sum Q_{\text{ground\_truth}}(s,a)} \times 100\%
    \end{equation}
    where $Q_{\text{model}}$ and $Q_{\text{ground\_truth}}$ represent predicted revenues from experiment and ground\_truth models respectively.

We verified that Return\% and \( r_s \)  correlate strongly (r = 0.93) and both have the same ordering as online Revenue, which shows the offline metrics have strong validity.
    
\end{itemize}

\paragraph{\textbf{Baselines}} We compare AWRQ-Mixer against several baseline models.
\begin{itemize}
    \item DQN \cite{DQN}: Employs a single deep network to approximate Q-values in high-dimensional state spaces. 
    \item DRQN \cite{DRQN}: Extends DQN with a recurrent mechanism to handle partial observability.  
    \item DDQN \cite{DDQN}: Decouples action selection from evaluation to mitigate Q-value overestimation.  
    \item Averaged Ensemble-DQN \cite{averagedDQN}: Aggregates multiple Q-networks to reduce variance and enhance stability. 
    \item REM \cite{REM}: Combines Q-networks through random convex mixtures for richer exploration.  
    \item VDN \cite{VDN}: Decomposes the global Q-value into a sum of per-agent Q-values for cooperative policies.  
    \item QMIX \cite{QMIX}: Employs a mixing network to ensure monotonic relationships among individual Q-values, thus improving coordination efficiency in multi-agent systems.
    \item Weighted QMIX \cite{WEIGHTED_QMIX}: Uses dynamic weighting in QMIX to emphasize heterogeneous agent contributions.
\end{itemize}

\paragraph{\textbf{Experiment Results}} We comprehensively analyze the real-world logs from the display advertising system. As shown in Table~\ref{tab:offline_results}, AWRQ-Mixer achieves significant improvements on both evaluation metrics. Compared to REM, the single-agent model used in RL-MPCA \cite{RL-MPCA}, AWRQ-Mixer boosts \(r_s\) by 3.4\% and Return\% by 8.0\%. Multi-agent methods generally outperform single-agent ones, underlining the benefits of collaborative modeling.


\begin{table}[h]
\centering
\caption{Comparison of offline evaluation results across different baseline models.}
\label{tab:offline_results}
\begin{tabular}{lll@{\hskip 1.2em}l}
\toprule
 & $r_s$ & Return\%  \\
\midrule
\textbf{Single-Agent} \\
\quad DQN & 0.859($\pm$0.025) & 82.59($\pm$4.22)  \\
\quad DDQN & 0.860($\pm$0.019) & 85.46($\pm$3.18)  \\
\quad DRQN & 0.862($\pm$0.023) & 86.82($\pm$3.67)  \\
\quad Averaged Ensemble-DQN & 0.870($\pm$0.015) & 87.48($\pm$2.79)  \\
\quad REM (RL-MPCA) & 0.881($\pm$0.014) & 89.47($\pm$2.63)  \\
\addlinespace
\textbf{Multi-Agent} \\
\quad VDN & 0.896($\pm$0.018) & 95.10($\pm$1.94)  \\
\quad Weighted QMIX & 0.900($\pm$0.015) & 95.65($\pm$1.68)  \\
\quad QMIX & 0.902($\pm$0.017) & 96.00($\pm$1.63)  \\
\quad \textbf{AWRQ-Mixer (MaRCA)} & 0.911($\pm$0.009) & 97.26($\pm$1.01)  \\
\hline
\end{tabular}
\end{table}

\paragraph{\textbf{Ablation Study}} We assess the contributions of three key components in AWRQ-Mixer by removing each one separately and comparing these variants to the full model. Table~\ref{tab:ablation_results} presents the results. The full model attains the highest $r_s$ (0.911) and Return\% (97.26\%), while converging in the fewest environment steps (1.1 M) and exhibiting the lowest gradient variance (0.18).  

\begin{table*}[!t]
\centering
\caption{Ablation study results demonstrating the impact of core innovations: adaptive weighting (AW), softplus-based monotonic constraints (SMC), and variance-guided credit assignment (VGCA).}
\label{tab:ablation_results}
\begin{tabular}{lrr@{\hskip 1.2em}rr}
\toprule
 & $r_s$ & Return\%  & Convergence (M steps)  & Gradient-variance  \\
\midrule
\quad \textbf{AWRQ-Mixer (MaRCA)} & 0.911 ($\pm$0.009) & 97.26 ($\pm$1.01) & 1.1 ($\pm$0.10) & 0.18 ($\pm$0.05) \\
\quad AWRQ-Mixer w/o AW & 0.910 ($\pm$0.010) &  97.17 ($\pm$1.10) & 1.4 ($\pm$0.20) & 0.54 ($\pm$0.19) \\
\quad AWRQ-Mixer w/o SMC & 0.908 ($\pm$0.009) & 96.92 ($\pm$1.18)  & 1.3 ($\pm$0.15) & 0.41 ($\pm$0.12) \\
\quad AWRQ-Mixer w/o VGCA & 0.906 ($\pm$0.011) &  96.71 ($\pm$1.37) & 1.7 ($\pm$0.20) & 0.27 ($\pm$0.08)  \\
\hline
\end{tabular}
\end{table*}

\begin{itemize}
    \item {AW}. Removing AW decreases $r_s$ and Return\% only marginally (–0.1\% and –0.09\%), while the gradient variance triples (0.54 vs 0.18) and the number of steps to reach the same validation loss increases by 27\%.  The larger variance indicates that, without the TD-error–driven head re-weighting, noisy or mis-calibrated heads exert disproportionate influence, producing erratic updates and slowing convergence.
    \item {SMC}. Replacing Softplus with an absolute-value operator leads to higher gradient variance (0.41) and a 20\% slower convergence, accompanied by a further drop in both $r_s$ and Return\%.  Although absolute value enforces monotonicity, its discontinuity at zero amplifies gradient fluctuations around the decision boundary. The measured variance confirms this analytic expectation and explains the observed degradation.
    \item {VGCA}. Eliminating VGCA yields the largest decline in ranking quality (–0.51\% $r_s$) and offline return (–0.55\%), together with the slowest convergence (1.7 M steps).  By reallocating TD-errors according to action variance, VGCA accelerates credit propagation in sparse-reward regions. Without VGCA, the agents require more samples to achieve the same validation criterion, even though the raw gradient variance remains moderate (0.27).
\end{itemize}

Across all metrics, each component contributes additively. Their combination yields a consistently more stable, accurate, and faster-converging learner.


\subsubsection{MPC-based Revenue-Cost Balancer Experiment Results}

\paragraph{\textbf{Evaluation Metrics}} We introduce two complementary metrics that capture distinct aspects of computation allocation: utilization rate $\mu$ and overutilization rate $\nu$.

\begin{itemize}
    \item Utilization Rate ($\mu$): This metric quantifies the effective usage of available resources. A higher $\mu$ indicates better computation resource utilization within capacity limits.
    \begin{equation}
        \mu = \frac{1}{T}\sum_{i=1}^{T} \frac{\min (\hat{C}_t, C_m)}{C_m}
    \end{equation}
    
    \item Overutilization Rate ($\nu$): This metric measures the risk of exceeding the computation budget. A lower $\nu$ indicates fewer stability violations and reduced risk of service degradation.
    \begin{equation}
        \nu = \frac{1}{T}\sum_{i=1}^{T} \frac{\max (\hat{C}_t, C_m) - C_m}{C_m}
    \end{equation}
\end{itemize}


Balancing \(\mu\) and \(\nu\) is crucial for ensuring efficiency and stability. While increasing \(\mu\) raises the risk of exceeding limits, reducing \(\nu\) leads to conservative resource utilization.

\begin{table*}[!t]
    \centering
    \addtocounter{table}{1}
    \caption{Online A/B test results comparing Static, RL-MPCA, MaRCA (Feedback-Based), and MaRCA (MPC-Based).}
    \label{tab:online_results}
    \begin{tabular}{lcccccccc}
        \hline
        & Revenue & GMV & Impressions & Clicks & ROI & CTR \\
        \hline
        Static & +0.00\% & +0.00\% & +0.00\% & +0.00\% & +0.00\% & +0.00\%  \\
        DCAF & +3.67\%($\pm$0.20\%) & +6.16\%($\pm$3.68\%) & +4.92\%($\pm$0.04\%) & +5.38\%($\pm$0.09\%)  & +2.40\%($\pm$3.69\%) & +0.69\%($\pm$0.10\%)\\
        RL-MPCA  & +12.16\%($\pm$0.28\%) & +13.78\%($\pm$4.03\%)  & +9.07\%($\pm$0.03\%) & +15.37\%($\pm$0.10\%)  & +0.55\%($\pm$4.04\%) & +4.85\%($\pm$0.10\%)  \\
        MaRCA-Feedback & +14.93\%($\pm$0.43\%) & +15.65\%($\pm$5.39\%) & +11.67\%($\pm$0.04\%) & +17.79\%($\pm$0.13\%) & +0.64\%($\pm$5.41\%) & +5.58\%($\pm$0.14\%)  \\
       \textbf{MaRCA-MPC} & \textbf{+16.67\%}($\pm$0.24\%) & \textbf{+18.18\%}($\pm$3.95\%) & +14.24\%($\pm$0.03\%) & +19.51\%($\pm$0.07\%) & +1.29\%($\pm$3.96\%) & +5.22\%($\pm$0.08\%)  \\
        \hline
    \end{tabular}
\end{table*}


\paragraph{\textbf{Offline Experiment Results}} We compared the performance of the feedback-based and MPC-based revenue-cost balancers. Table \ref{ladm-performance} shows that the MPC-based approach not only improves overall computation resource usage but also substantially lowers the risk of exceeding the computation budget.

\begin{table}[h]
\centering
\addtocounter{table}{-2}
\caption{Comparison of load utilization and overutilization rates between feedback-based and MPC-based revenue-cost balancer.}
\begin{tabular}{cccc}
\hline
Method & Utilization Rate & Overutilization Rate \\
\hline
Feedback-based & 92.33\%($\pm$0.40\%) & 2.91\%($\pm$0.43\%)  \\
\textbf{MPC-based} & 95.29\% ($\pm$0.23\%) & 0.64\%($\pm$0.10\%) \\
\hline
\end{tabular}
\label{ladm-performance}
\end{table}

\paragraph{\textbf{Hyperparameter Analysis}} We investigate three key hyperparameters: decay-weighting factor $\alpha$, oscillation damping factor $\beta$, and prediction horizon $N$.

\begin{itemize}
\item {Decay-Weighting Factor} $\mathbf{\alpha}$. A small \(\alpha\) prioritizes current returns, which can lead to more aggressive resource utilization but can intensify fluctuations. Conversely, a large \(\alpha\) may cause underutilization by overemphasizing future risks. As shown in Figure~\ref{hp-analysis1}, the overutilization rate stabilizes around \(\alpha=0.4\), indicating balanced short/long-term trade-offs. We choose $\alpha = 0.4$ as the parameter value. 

\begin{figure}[ht]
  \centering
  \begin{subfigure}[b]{0.28\textwidth}
    \includegraphics[width=\textwidth]{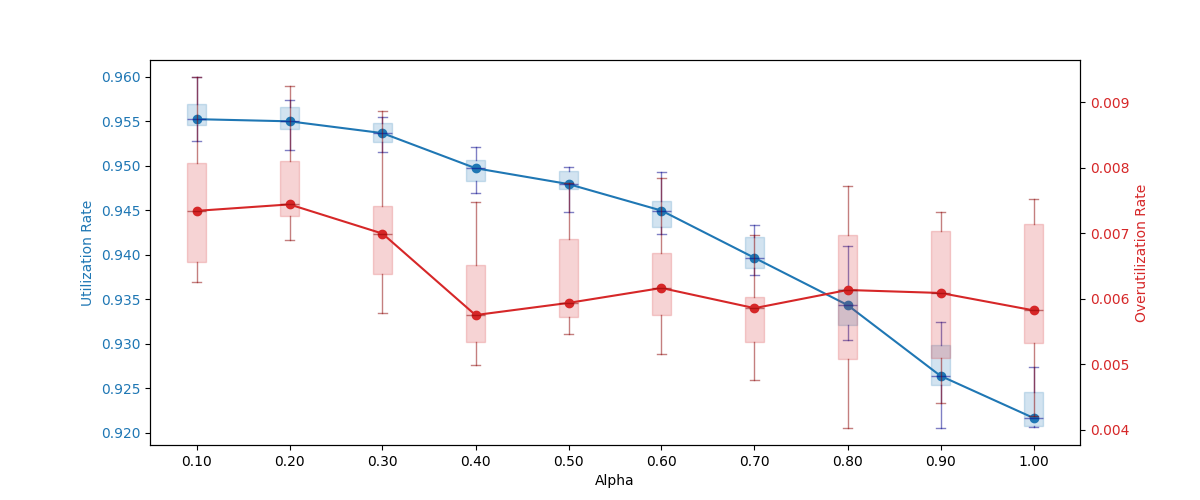}
    \caption{Decay‐Weighting Factor $\alpha$}
    \label{hp-analysis1}
  \end{subfigure}
  \hfill
  \begin{subfigure}[b]{0.28\textwidth}
    \includegraphics[width=\textwidth]{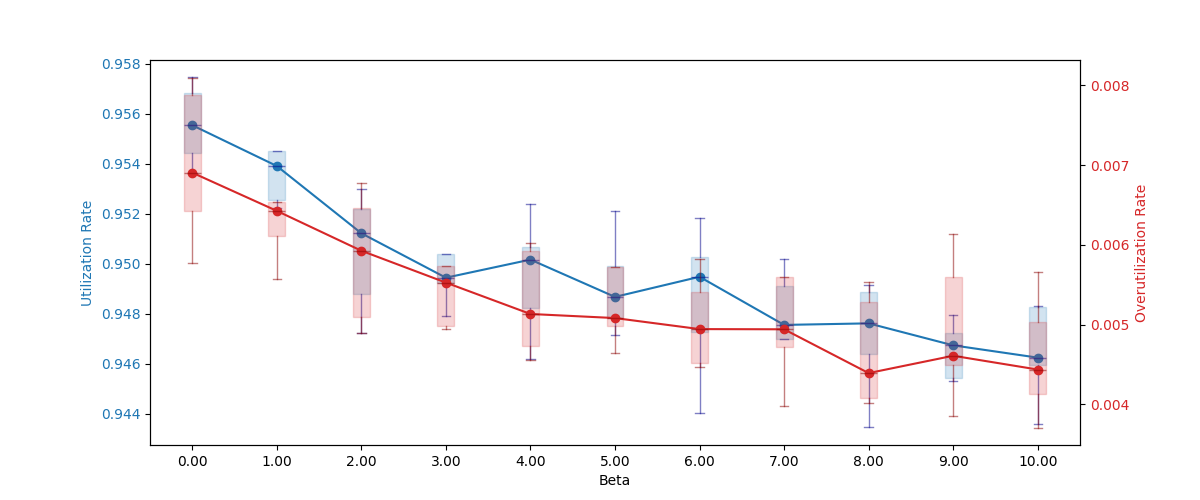}
    \caption{Oscillation Damping Factor $\beta$}
    \label{hp-analysis2}
  \end{subfigure}
  \hfill
  \begin{subfigure}[b]{0.28\textwidth}
    \includegraphics[width=\textwidth]{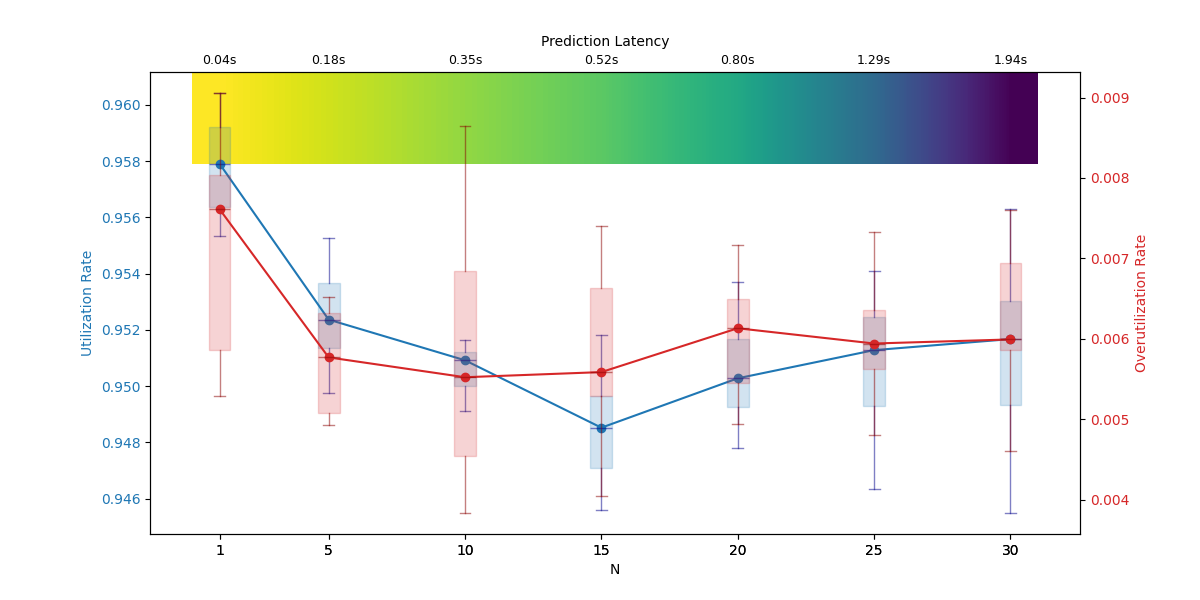}
    \caption{Prediction Horizon $N$}
    \label{hp-analysis3}
  \end{subfigure}

  \caption{Hyperparameter analysis results for $\alpha$, $\beta$ and $N$.}
  \label{fig:hp-analysis-all}
\end{figure}


\item {Oscillation Damping Factor} $\mathbf{\beta}$. $\beta$ is employed to limit the fluctuation of computation resource utilization. When $\beta$ is too high, the system becomes overly conservative, lowering utilization. Conversely, an extremely small $\beta$ can result in unstable spikes. Based on the results in Figure~\ref{hp-analysis2}, we choose \(\beta = 8\) to achieve a trade-off between resource utilization and stability.


\item {Prediction Horizon} \(\mathbf{N}\). While a larger \(N\) often improves long-term optimization, it simultaneously increases forecasting errors and latency overhead. Conversely, a smaller \(N\) may fail to capture future dynamics. Figure~\ref{hp-analysis3} indicates that \(N = 10\) provides an effective balance between predictive accuracy and runtime efficiency.

\end{itemize}

\subsection{Online A/B test Results} 
\label{onlineabtestresults}

We conducted a four-week online A/B test comparing five strategies: a static method, DCAF \cite{DCAF}, RL-MPCA \cite{RL-MPCA}, MaRCA with a feedback-based Revenue-Cost Balancer, and MaRCA with an MPC-based balancer. The static approach uses fixed allocation rules through stress testing and practical experience, with predefined downgrades to manage traffic spikes. Table \ref{tab:online_results} reports impressions, clicks, revenue, gross merchandise volume (GMV), return on investment (ROI), click-through rate (CTR), and cost per mille (CPM), with revenue and GMV as our primary metrics. Here, ROI is defined as $\mathrm{ROI}=\mathrm{GMV}/\mathrm{Spend}$, where $\mathrm{Spend}$ denotes advertiser spend (i.e., the platform's revenue in our setting). Our near-real-time deployment adds virtually no additional latency.



MaRCA achieved statistically significant improvements across all key metrics while operating within existing computation resource constraints. After the four-week A/B test, MaRCA was rolled out to 100\% of production traffic, where it now processes hundreds of billions of requests each day and has since supported multiple large‑scale sales events. Its stability has also mitigated the need for continuous on-call support.

\section{Conclusion}
In this paper, we propose MaRCA to address the challenge of maximizing business revenue in large-scale recommender systems under computation resource constraints. By modeling recommendation stages as cooperative agents and integrating Centralized Training with Decentralized Execution, MaRCA effectively captures cross-stage dependencies while preserving independent decision-making. Additionally, we introduce an MPC-based revenue-cost balancer that proactively adjusts resource allocation, ensuring system stability under dynamic traffic conditions. Our extensive offline experiments and large-scale online deployment demonstrate that MaRCA significantly improves business revenue, achieving a 16.67\% revenue increase with no additional computation resource. Future work may focus on enhancing model capabilities, expanding the action space, and exploring cross-domain applications. 

\bibliographystyle{ACM-Reference-Format}
\bibliography{sample-base}

\clearpage

\appendix

\section{Lagrangian Relaxation Details}
\label{sec:lagrangian}

To satisfy the budget constraint in Eqs. \eqref{eq:constr1}--\eqref{eq:constr3} while maximizing total revenue in Eq. \eqref{eq:obj}, we adopt a Lagrangian relaxation approach:

\begin{equation}
 \label{eq:lagrangian}
\mathcal{L} = \sum_{i=1}^M \sum_{\mathbf{a} \in \mathcal{A}} x_{i, \mathbf{a}}\,Q(s_t,\mathbf{a}) 
- 
\lambda_t  \,\sum_{i=1}^M \sum_{\mathbf{a} \in \mathcal{A}} (x_{i,\mathbf{a}}\,C(s_t, \mathbf{a}) - C_m))
\end{equation}

Where Revenue-Cost Balancer (\(\lambda_t\)) serves as a trade-off parameter balancing business revenue and computation cost.  The constant term $\lambda_t C_m$ can be omitted in the maximization as it does not affect the optimization. Substituting the Lagrangian utility in Eq. \eqref{eq:lagrangian} back into the original objective of Eq. \eqref{eq:obj} absorbs the cost constraint into the multiplier $\lambda_t$. The global problem becomes:

\begin{equation}
\max_{x_{i,\mathbf{a}}} \sum_{i=1}^{M} \sum_{\mathbf{a} \in \mathcal{A}} x_{i,\mathbf{a}}\bigl(Q(s_t,\mathbf{a}) - \lambda_t C(s_t,\mathbf{a})\bigr)
\end{equation}

Because $x_{i,\mathbf{a}}$ is one‑hot for each request, the double sum decomposes into independent per‑request arg‑max operations. Following the approach in \cite{RL-MPCA}, each user request effectively solves a local decision subproblem:

\begin{equation}
\mathbf{a}^* = \arg\max_{\mathbf{a}\in\mathcal{A}} \left( Q(s_{t}, \mathbf{a}) - \lambda C(s_{t}, \mathbf{a}) \right)
\end{equation}

which is Eq.~\eqref{eq:a_star} in the main text.

\section{AutoBucket TestBench}
\label{sec:autobucket}

Accurately estimating computation costs is essential for resource allocation in large-scale recommender systems. However, real-world cost labels are scarce, and the variability in action results across multiple stages complicates direct cost measurement at the request level. We propose a two-phase approach that predicts action results and derives their associated costs via cost mapping. This method leverages the observation that computation costs remain consistent for identical action results, even when the underlying requests differ. Since computation cost is primarily dictated by code logic, model complexity, and retrieval processes.

\begin{figure}[ht]
    \centering
    \includegraphics[width=\linewidth]{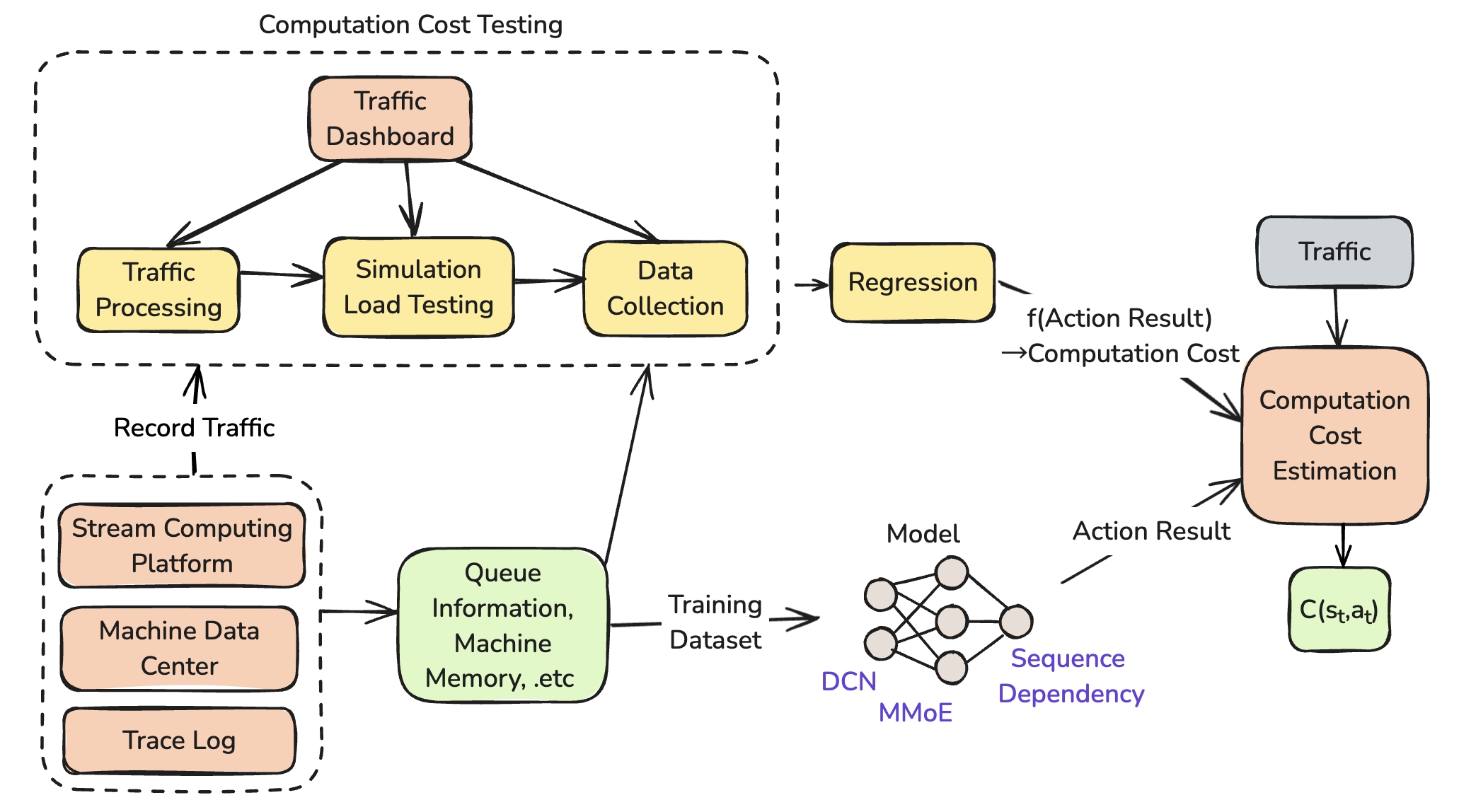}
    \caption{AutoBucket TestBench enables multi-stage computation cost estimation through traffic simulation, regression analysis, and sequence-aware modeling.}
    \Description[Alt Text]{Diagram illustrating the AutoBucket TestBench workflow, which includes traffic simulation, regression analysis, and computation cost estimation.}
    \label{fig:autobucket_architecture}
\end{figure}

Figure~\ref{fig:autobucket_architecture} presents the core architecture of our AutoBucket TestBench. Traffic logs from multiple sources are processed and subjected to simulated load testing, during which computation resource utilization is recorded. The resulting labeled samples (e.g., queue length vs.\ cost) are aggregated and used to fit a regression model, producing a cost-mapping function. Combined with predicted action outcomes, this module enables accurate computation cost estimations.

\noindent \textbf{Action Results Prediction}. Given user and contextual features, we employ a DCN \cite{DCN}+MMoE \cite{mmoe} hybrid architecture to predict action results. The Deep \& Cross Network (DCN) captures nonlinear feature interactions, while the Multi-gate Mixture of Experts (MMoE) layer enables parameter sharing among predictions. This multi-task learning framework operates within the $S \times A$ space, leveraging abundant logged interaction data for supervised training.

\noindent \textbf{Computation Cost Estimation}. With predicted action results, we estimate the corresponding computation cost via a cost mapping.

\begin{itemize}
    \item \textbf{Elastic Queue}. For actions determining queue truncation lengths, we use empirical tests under varied queue lengths to measure computation cost, and then fit a regression model. 
        \begin{enumerate}
            \item \textbf{AutoBucket}. Simulation traffic is grouped into buckets based on request value and queue length to capture the variability in computation cost across diverse traffic conditions.
            \item \textbf{TestBench}. The TestBench processes requests in each bucket at a fixed queries-per-second (QPS).
            The TestBench is deployed on \(n\) machines, each with \(N_{\text{cores}}\) computation resource cores, and the computation resource utilization \(p\%\) is recorded during these tests. The computation cost per request for each bucket is computed as:
                \begin{equation}
                   \text{Computation Cost per Bucket} = \frac{p\% \times n \times N_{\text{cores}}}{QPS}
                \end{equation}
            A monotonic regression model is then used to fit the relationship between queue lengths and computation cost. 
        \end{enumerate}
        
    \item \textbf{Elastic Model}. For actions that involve switching between different models, computation costs are derived from independent measurements of each switch configuration. Similar to Elastic Queue, Elastic Models are tested under controlled conditions to measure their computation cost. 
    \item \textbf{Elastic Channel}. The computation cost is derived as the cumulative sum of the computation costs across all selected actions.
\end{itemize}

\section{Hyperparameters Sensitivity Analysis}
\label{sec:hyperparameters-sens}
We conduct an offline sensitivity analysis for two key hyperparameters in AWRQ-Mixer: discount factor $\gamma$ (see Table~\ref{tab:hp_gamma}) and the number of ensemble heads $K$ (see Table~\ref{tab:hp_heads}). Each configuration is evaluated over 5 random seeds. 

\begin{itemize}
\item {Discount Factor} $\gamma$. We selected a $\gamma$ value of 0.9. In reinforcement learning, $\gamma$ represents the trade-off between long-term and immediate rewards, determining how much importance the model places on future rewards. From Table~\ref{tab:hp_gamma}, we observe that when $\gamma$ is set to 0.90, the model exhibits the highest stability and performance. In contrast, other values such as 0.70 and 0.99, show slightly lower results. Therefore, we ultimately chose 0.9 for $\gamma$ to ensure a balance of strong performance and system stability.

\begin{table}[h]
  \centering
  \caption{Discount Factor \(\gamma\) Sensitivity }
  \label{tab:hp_gamma}
  \begin{tabular}{lcc}
    \toprule
    $\gamma$ & $r_s$     & Return\% \\
    \midrule
    0.70 & 0.9096(\(\pm\)0.0076) & 97.18(\(\pm\)0.97)  \\
    0.80 & 0.9107(\(\pm\)0.0083) & 97.22(\(\pm\)1.01)  \\
    0.90 & 0.9112(\(\pm\)0.0086) & 97.26(\(\pm\)1.01)  \\
    0.99 & 0.9105(\(\pm\)0.0109) & 97.21(\(\pm\)1.05)  \\
    \bottomrule
  \end{tabular}
\end{table}

\item {Number of ensemble heads} $K$. Table~\ref{tab:hp_heads} illustrates the effects of different ensemble head numbers on performance. As the value of $K$ increases, we see a gradual improvement in both $r_s$ and Return\%. Specifically, when $K$ is set to 200, the $r_s$ value reaches $0.9120$, with a Return\% of $97.30$. However, while $K$ = 200 yields the best performance, increasing the number of ensemble heads significantly boosts computational complexity. After considering the trade-off between computational resources and model effectiveness, we selected $K$ = 20, which provided nearly optimal results while minimizing unnecessary computational overhead.
\end{itemize}


\begin{table}[h]
  \centering
  \caption{Ensemble Head \(K\) Sensitivity}
  \label{tab:hp_heads}
  \begin{tabular}{lcc}
    \toprule
    $K$ & $r_s$     & Return\% \\
    \midrule
    1   & 0.9065(\(\pm\)0.0095) & 96.78(\(\pm\)1.11)  \\
    5   & 0.9081(\(\pm\)0.0089) & 96.95(\(\pm\)1.02)  \\
    20  & 0.9112(\(\pm\)0.0086) & 97.26(\(\pm\)1.01)  \\
    100 & 0.9118(\(\pm\)0.0087) & 97.29(\(\pm\)1.01)  \\
    200 & 0.9120(\(\pm\)0.0085) & 97.30(\(\pm\)1.00)  \\
    \bottomrule
  \end{tabular}
\end{table}

\section{Hyperparameters}
\label{sec:hyperparameters}
The following Table \ref{tab:awrq_hyperparams} lists the hyperparameters we used in experiments. 

\begin{table}[h]
\centering
\caption{Hyperparameter settings for MaRCA.}
\label{tab:awrq_hyperparams}
\begin{tabular}{lll@{\hskip 1.2em}}
\toprule
 Hyperparameters & Value  \\
\midrule
\textbf{AWRQ-Mixer} \\
\quad Learning rate & 0.01  \\
\quad GRU hidden size & 256  \\
\quad Discount factor $\gamma$ & 0.9  \\
\quad Target-network update frequency $\tau$ & 100  \\
\quad Ensemble size $K$ & 20  \\
\quad $\varepsilon$-greedy exploration rate & 0.05  \\
\quad Size of hidden layer in the network & [512, 256]  \\
\quad Optimizer & Adam  \\
\quad Dropout rate & 0.2  \\
\quad Weight initializer & glorot uniform  \\
\quad Batch size & 2048  \\
\addlinespace
\textbf{MPC-based Revenue-Cost Balancer} \\
\quad Decay-Weighting Factor $\alpha$ & 0.4  \\
\quad Oscillation Damping Factor $\beta$ & 8  \\
\quad Prediction Horizon $N$ & 10  \\
\hline
\end{tabular}
\end{table}


\section{Online Serving of AWRQ–Mixer}
Algorithm \ref{alg:awrqmixer-serving} shows AWRQ–Mixer's online serving process for a given request.

\begin{algorithm}[ht]
\caption{Online Serving of AWRQ–Mixer }
\label{alg:awrqmixer-serving}
\begin{algorithmic}[1]
\REQUIRE Trained networks $\{Q_{g}^{k}\}$.

\FOR{each incoming request}
    \STATE Observe state features $o_t$
    \STATE AWRQ outputs $K$ Q-values $ Q_{g}^k(o_t, a_t)$
    \STATE $Q_{{g}} \leftarrow \frac{1}{K}\sum_{k=1}^{K} Q_{g}^{k}$  
\ENDFOR
\end{algorithmic}
\end{algorithm}

\section{Online Serving of MaRCA}
Algorithm \ref{alg:marca-serving} shows MaRCA's online serving process for a given request.

\begin{algorithm}[ht]
\caption{Online Serving of MaRCA}
\label{alg:marca-serving}
\begin{algorithmic}[1]
\REQUIRE Trained networks $\{Q_g\}$, cost estimator $C(\cdot)$,  
         MPC Revenue-Cost Balancer that outputs $\lambda_t$
\STATE {\raggedright \textbf{if} periodic update moment, $\lambda_t \leftarrow \text{MPC Revenue-Cost Balancer} $\par}

\FOR{each incoming request}
    \STATE Observe state $s_t$
    \STATE $\mathbf{a}^* \leftarrow \operatorname*{arg\,max}_{\mathbf{a}\in\mathcal{A}} \left( Q(s_{t}, \mathbf{a}) - \lambda C(s_{t}, \mathbf{a}) \right)$
    \STATE Execute joint action $\mathbf a^{*}$
\ENDFOR
\end{algorithmic}
\end{algorithm}

\end{document}